\documentstyle[aps,tighten,graphicx,prl,twocolumn]{revtex}

\begin{document}
\title{The Einstein-Podolsky-Rosen 
Paradox and Entanglement 1: Signatures of EPR
correlations for continuous variables
\\}
 \vskip 1 truecm
\author{M. D. Reid\\}
\address{Physics Department, 
The University of Queensland,Brisbane,
Australia}
\date{\today}
\maketitle
\vskip 1 truecm
\begin{abstract}
 A generalization of the 1935 Einstein-Podolsky-Rosen (EPR) argument for
measurements with continuous variable outcomes is presented to establish
criteria for the demonstration of the EPR paradox, for situations where the
correlation between spatially separated subsystems is not perfect. Two types of
criteria for EPR correlations are determined. The first type are based on
measurements of the variances of conditional probability distributions and are
necessary to reflect directly the situation of the original EPR paradox. The
second weaker set of EPR criteria are based on the proven failure of
(Bell-type) local realistic theories which could be consistent with a local
quantum description for each subsystem. The relationship with criteria
sufficient to prove entanglement is established, to show that any demonstration
of EPR correlations will also signify entanglement. It is also shown how a
demonstration of entanglement between two spatially separated subsystems, if
not able to interpreted as a violation of a Bell-type inequality, may be
interpreted as a demonstration of the EPR correlations. In particular it is
explained how the experimental observation of two-mode squeezing using
spatially separated detectors will signify not only entanglement but EPR
correlations defined in a general sense. 
\end{abstract}
\narrowtext
\vskip 0.5 truecm
\section{Introduction}

In 1935 Einstein, Podolsky and 
Rosen~\cite{epr} (EPR) presented their famous argument in an attempt to 
show that quantum mechanics is incomplete. EPR 
defined a premise called local realism, which is assumed in all 
classical theories. 
 The premise of realism implies 
 that if one can predict with 
 certainty the result of a measurement of a physical quantity 
 at a location $A$, without disturbing the 
 system at $A$, then the results of the measurement are 
 predetermined. In these circumstances there is an 
 ``element of reality'' corresponding to this 
 physical quantity, the  element of reality being a variable that assumes 
 one of the set of values that are the predicted results of the 
 measurement. 
  The locality (no action-at-a-distance) 
 assumption implies that a measurement performed
at a spatially separated location $B$ cannot induce any immediate 
change to the 
subsystem at $A$.
 Local realism is defined as the dual premise, where 
EPR's  
realism and locality  are both assumed.

 For certain correlated 
spatially separated particles EPR showed that, 
if quantum mechanics is to be consistent
 with local 
realism,
the position and momentum of a 
single localised particle must be simultaneously predetermined with 
absolute definiteness. Such 
simultaneous determinacy for both position and momentum is not 
predicted for any quantum state.
 While EPR took the view 
 that local realism must be valid and therefore argued that      
 quantum mechanics was incomplete, the argument is perhaps best viewed as a 
 demonstration of the inconsistency between quantum mechanics as we 
 know it (that is without ``completion'') and local 
 realism. 
 
 Observation of an EPR correlation, while not as conclusive as 
 observing a violation of a Bell inequality~\cite{Bell,CH}, can be 
 carried out at high detector efficiency.
  It is possible to measure continuous variable EPR correlations 
  using fields, where the 
 conjugate ``position'' and 
 ``momentum'' observables are replaced by   
 the two orthogonal noncommuting 
 quadrature phase amplitudes of the field~\cite{eprquad}. These field 
 quadrature amplitudes are measurable using homodyne detection schemes.
    A specific  criterion, a violation of an inferred Heisenberg 
    Uncertainty Principle (H. U. P.), to 
 demonstrate the continuous variable 
 EPR paradox for real experiments
  was put forward in 1989~\cite{eprr}. This proposal employed a 
 two-mode squeezed state~\cite{cavessch} as the source.

 The first experimental achievement, using the parametric 
 oscillator~\cite{rd}, of 
 the 1989 EPR criterion, for 
   efficient measurements with 
   continuous variable outcomes,
    was reported by Ou et al~\cite{ou}  
 in 1992. Recently
  Zhang et al~\cite{zhang} detected EPR correlations between the  
the intense output fields of the 
parametric oscillator above threshold. Silberhorn et al~\cite{silber}
 have detected such 
correlations for 
pulsed fields, and there have been further theoretical 
proposals~\cite{eprtaraag,eprtomb}.
 The EPR fields have 
   proven significant in enabling the experimental 
  realization of a
   continuous variable quantum teleportation~\cite{tele}, 
 and may have application also to continuous variable 
 quantum cryptography~\cite{crycont}.

It is known  that for the two-mode squeezed state which  exhibits the EPR 
paradox for continuous variable quadrature amplitude 
measurements, a violation of a 
Bell inequality~\cite{Bell,CH,Belltwo} is not possible, for the same sorts of 
measurements.  This is transparent~\cite{Bell} on recognizing that the Wigner 
function is positive for such a state and can act as a local hidden variable theory 
for the predictions of such measurements. It is for this reason that 
the Bell inequality must be replaced with an EPR or entanglement criterion in 
certain quantum cryptographic or teleportation schemes involving 
continuous variable measurements. 

In view of this, in order to understand the application of the 
continuous variable EPR correlations 
 to areas of research  such as quantum cryptography, and to assist in required 
 proofs of security, our first objective in this first paper is to fully   
{\it define  EPR correlations 
 through a generalization of the EPR paradox} (rather than through 
 violations of Bell inequalities) for spatially separated 
 systems where the  correlation need not be maximum.  
In Sections 2 and 3 we review and further develop the 
generalization of the EPR argument to provide criteria for 
demonstrating EPR correlations  
 for situations where the 
correlation between spatially separated subsystems is less than 
optimal.  To enable experimental proof of EPR correlations, the 
criteria are to be expressed in terms of measurable 
probabilities and correlations. 

Two types of EPR criteria are established. The first, a stronger 
set (Sections 3a,b and c), are based on reduced variances of conditional probability 
distributions, and represent most directly the application of EPR's 
premises of realism and no action-at-a-distance and are therefore in 
the spirit of the original EPR paradox.
The second  are a weaker set, established indirectly: through the failure of any  
 local realistic theory which is also consistent with a local quantum 
description for each subsystem, to adequately predict the quantum statistics.

Such EPR criteria are of a {\it fundamental significance in providing the 
condition for the experimental demonstration of the EPR paradox}. 
 The point is made, as has been made previously~\cite{Bell}, that the 
 demonstration of EPR is not itself be a demonstration of the 
failure of all local hidden variable (or local realistic) theories. 
It is however a failure of all such theories that could be consistent 
with each of the spatially separated subsystems being depicted by a quantum 
state, and as such is a {\it confirmation of quantum inseparability 
(entanglement~\cite{sch}) for the striking situation of spatially separated 
subsystems}. 
 The EPR criteria define conditions for the onset of a 
fundamental philosophical conflict between the viewpoint of a believer 
in local realism, and the believer that the quantum state 
 represents the ultimate (most 
complete)  
description of a physical system. 

It may be the case that further different types of 
measurements will show outright the failure of all local hidden 
variable theories, by way of violation of a Bell inequality. 
It is still relevant 
however to {\it recall the significance of the EPR demonstrations 
alone},   
 for the following reasons. First, such demonstrations have actually
been performed with reasonable spatial separations (though not with 
causally separated events) for  high 
detection efficiencies. This means that the conclusions are loop-hole 
free in that they do not require the auxiliary assumptions 
associated with the category of Bell inequality tests, that have been 
performed using as subsystems fields with significant 
spatial separations, but inefficient detectors~\cite{Belltwo} or vice 
versa~\cite{wine}. 

Second, the EPR 
correlations have been proven experimentally for macroscopic systems, 
and also where 
measurement outcomes are continuous and can be translated with some 
minor experimental modification to a 
macroscopic outcome domain~\cite{mdrmacro}. This is not the 
case for Bell tests which are difficult in such 
regimes~\cite{bellmacronum}. 
 For the sake of the completeness of this paper then the  
{\it philosophy and significance of the EPR argument and the 
significance of its demonstration} is revised in Section 
4.

In Section 5 the link between  a demonstration 
of entanglement and a demonstration of the EPR paradox is 
established.  Here we define a demonstration of 
entanglement as an experimentally measured 
violation of a necessary criterion 
for separability, where the criterion must be expressed in terms of 
measurable probabilities or correlations in the fashion of a 
Bell-inequality. Such entanglement criteria have been derived 
 for measurements with continuous variable outcomes by Duan et 
al~\cite{content} and Simon~\cite{content}. We prove explicitly 
an entanglement criterion which is based on the achievement of 
two-mode squeezing between certain spatially separated quadrature 
amplitudes.
It is discussed then how such a suitably performed demonstration 
of entanglement will also demonstrate the failure of a weaker EPR criterion. 
It is also proved explicitly that the 
demonstration of an EPR paradox must always imply a demonstration of 
entanglement: the EPR criteria are signatures of entanglement. 
The  results of this paper have been 
presented in brief in  
a previous preprint~\cite{qpr} 
(quant-ph 0103142), but are derived here 
  in full detail.

The  continuous variable 
 measurements 
that demonstrate the EPR paradox for the two-mode squeezed state 
are not predicted to directly show a 
violation of a Bell inequality. However  for certain quantum 
cryptographic protocols one may 
replace the Bell-inequality  
 by 
an EPR-criterion to test for security.  A proof of security for the 
scheme similar to that proposed by Ekert~\cite{numbercry}, but based 
on the use of 
EPR criteria, rather than Bell inequalities, is presented in a  
second paper~\cite{mdrcrytwo}.  

 \section{Generalization of the EPR argument}
 
 In order to demonstrate the existence of continuous variable 
 EPR correlations for 
real experiments, we need to extend the 
EPR argument to situations where the results of measurements between 
the spatially separated subsystems need not 
be maximally correlated. This is also necessary if we are to link  
EPR criteria with entanglement criteria, since it is certainly 
true that not all 
entangled states are perfectly correlated.

    EPR originally argued as follows. Consider two spatially separated subsystems 
   at $A$ and $B$. EPR considered two 
 observables $\hat{x}$ (the ``position'') and $\hat{p}$ (``momentum'') for subsystem $A$, where $\hat{x}$ 
 and $\hat{p}$ do not commute, so that ($C$ is nonzero) 
 \begin{equation}
 [\hat{x},\hat{p}]=2C.
 \end{equation}
 Suppose one may predict with certainty the result of 
 measurement $\hat{x}$, based on the result of a measurement performed at $B$. 
  Also, 
  for a different choice of measurement at $B$, suppose one may  predict the 
 result of measurement  $\hat{p}$ at $A$. Such correlated systems are 
 predicted by quantum theory. Assuming ``local realism'' (discussed in 
 Section 1)
  EPR  
 deduce the existence 
 of an ``element of reality'', ${\tilde x}$, for 
 the physical quantity $\hat{x}$; 
 and also 
 an element of 
 reality, ${\tilde p}$, for $\hat{p}$. 
   Local realism implies the 
  existence of two hidden variables ${\tilde x}$ and ${\tilde p}$ that 
  simultaneously 
  predetermine, with no uncertainty, the values for the result of an 
  $\hat{x}$ or $\hat{p}$ measurement on subsystem $A$, should it be performed. 
  This 
  hidden variable state for the subsystem $A$ alone is not describable within 
  quantum mechanics, since simultaneous eigenstates of $\hat{x}$ 
  and $\hat{p}$ do not exist.
   Hence, EPR argued, if quantum mechanics 
    is to be compatible with local realism, we must regard quantum 
    mechanics to be incomplete. 
        
 We now need to extend the 
EPR argument to situations where the result of measurement $\hat{x}$ at $A$ 
cannot be predicted with absolute certainty~\cite{eprr,qpr}.       
 The assumption of local 
  realism allows us to deduce the existence of an ``element 
  of reality'' of some type for $\hat{x}$ at $A$,
   since we can make a prediction of the 
  result at $A$, without disturbing the subsystem at $A$, under the locality 
   assumption. This prediction is subject to the result $x_{i}^{B}¥$ of a 
   measurement, $\hat{x}^{B}$ say, performed 
   at $B$. (Throughout this paper $i$ is used to label the possible 
   results, discrete or otherwise, of the measurement $\hat{x}^{B}$).
  The predicted results for the 
  measurement at $A$, 
  based on the measurement at $B$,  
  are however no longer a set of definite 
  numbers with zero uncertainty, but become fuzzy, being described by  a set 
  of distributions $P(x|x_{i}^{B}¥)$ giving the probability of a result 
  for the
    measurement at $A$, conditional on a result $x_{i}^{B}¥$ for measurement 
    at $B$. We define  $\Delta_{i}^{2}x$ to be the variance of the 
    conditional distribution $P(x|x_{i}^{B}¥)$.

  Similarly we may infer the result of  
 measurement $\hat{p}$ at $A$, based on a (different) measurement,
  $\hat{p}^{B}¥$ say, at $B$. Denoting the results of the measurement 
  $\hat{p}^{B}¥$ 
  at $B$ by $p_{j}^{B}¥$, we then define the 
 probability distribution,  $P(p|p_{j}^{B}¥)$ which is the predicted result of 
 the measurement for $\hat{p}$ at $A$ 
conditional on the result $p_{j}^{B}¥$ 
 for the measurement $\hat{p}^{B}¥$ at $B$. 
   The variance of the conditional  distribution $P(p|p_{j}^{B})$ 
  is denoted by $\Delta_{j}^{2}p$.

\section{Signatures of the EPR paradox}

  For a given experiment one could in principle measure the individual 
variances $\Delta_{i}^{2}¥x$  of the conditional 
distributions $P(x|x_{i}^{B}¥)$ (and 
also $\Delta_{j}^{2}¥p$ for the $P(p|p_{j}^{B}¥)$). 
 If each of the variances satisfy 
 \begin{eqnarray}
 \Delta_{i}^{2}¥x&=&0\nonumber\\ 
\Delta_{j}^{2}¥p&=&0
\end{eqnarray}
 (for all $i,j$) of course there is no difficulty 
in establishing that this would imply the demonstration of the 
original EPR 
paradox. 

This situation however is not practical for 
continuous variable measurements. Instead of considering the problem 
of simultaneous eigenstates as originally proposed by EPR, we suggest 
instead a different and experimentally realizable criterion based on 
the Heisenberg Uncertainty Principle (H. U. P.)
\begin{equation}
\Delta \hat{x} \Delta \hat{p}\geq C \label{eqn:hup}
\end{equation}
For the sake of notational convenience we now consider in the remainder of 
the paper that  
appropriate scaling enables $\hat{x}$ and $\hat{p}$ to be 
dimensionless and $C=1$.

\subsection{Strong EPR correlations using bounded conditional 
distributions}

EPR correlations however would be demonstrated in a convincing 
manner if 
the experimentalist could
 measure each of the conditional distributions $P(x|x_{i}^{B}¥)$
  and establish that each of the distributions is very narrow, 
 in fact  constrained  
so that 
\begin{eqnarray}
P(x|x_{i}^{B}¥)&=&0 \quad,if\quad|x-\mu_{i}|>\delta \nonumber\\
P(p|p_{j}^{B}¥)&=&0 \quad,if\quad|p-\nu_{j}|>\delta
\end{eqnarray}
 Here $\mu_{i}$ is the mean of the conditional 
distribution $P(x|x_{i}^{B})$ and $\nu_{j}$ is the mean of the conditional 
distribution $P(p|p_{j}^{B})$.

In this case the assumption of local realism would 
imply, since the measurement $\hat{x}^{B}$ at $B$ will {\it always} imply 
the result of $\hat{x}$ at $A$ to be within the range $\mu_{i}\pm\delta_{x}¥$, 
that the 
result of the measurement at $A$ is predetermined to be within a 
bounded 
range of width $2\delta$. 
 In a straightforward extension of EPR's argument, we 
replace the words ``predict with certainty'' with ``predict 
with certainty that the result is constrained to be within the range  
 $\mu_{i}¥\pm\delta$'', and then 
 define an ``element of reality'' with this intrinsic bounded 
blurring or fuzziness $\delta$. 

After considering the 
$\hat{p}$ and $\hat{p}^{B}$ correlations, and where 
$\delta<1$, the conclusions of the paradox 
follow. This is because the predetermined precision associated with the 
``elements of reality'' describing the subsystem at $A$ 
could not be given by any quantum state, in view of the uncertainty 
relation (\ref{eqn:hup}). This situation represents the spirit of the original EPR 
gedanken experiment in its truest form. Such narrow  
distributions over the entire outcome domain 
$x_{i}^{B}$ (and $p_{j}^{B}$) 
 have however to my knowledge not been 
experimentally established.  In the case of the two-mode squeezed 
state (\ref{eqn:twomode}) to be discussed later, this situation is never 
strictly mathematically satisfied for finite values of the squeezing parameter
$r$ as defined in equation (\ref{eqn:twomode}). It may be satisfied 
to a high degree of certainty, but this  
requires  values for the squeezing parameter $r$ 
 that are difficult to achieve experimentally.

  \subsection{1989 EPR criterion: Violation of an Inferred Heisenberg Uncertainty 
 Principle}

 We next consider the situation where an experimenter has 
 demonstrated 
  that for 
 {\it every outcome} $x_{i}^{B}$ (and $p_{j}^{B}$) 
 for the  measurement $\hat{x}^{B}$ 
(and $\hat{p}^{B}$) performed at $B$, the variance $\Delta_{i}x$ 
(and $\Delta_{j}p$) of the appropriate 
conditional distribution satisfies 
\begin{eqnarray}
\Delta_{i}x&<&1\nonumber\\
\Delta_{j}p&<&1\label{eqn:eprcondeq}
\end{eqnarray}
for all $i,j$. The measurement at $B$ always allows an inference 
of the result at $A$ to a precision better than given by the uncertainty 
bound $1$. 

In this case we do not predict a result at $A$ ``with 
certainty'', as in EPR's original paradox. The measurement 
$\hat{x}^{B}$ at $B$ 
however does predict with a certain probability constraints on 
the result for $\hat{x}$ at $A$. Following the EPR argument, which assumes no 
action-at-a-distance, so that the 
measurement at $B$ does not cause any instantaneous 
influence to the system at $A$, 
one can attribute a probabilistic predetermined
 ``element of reality'' to the system at $A$. 
 There is a similar predicted result for the measurement $\hat{p}$ 
 at $A$ based on a result of measurement at $B$, and a corresponding 
 predetermined description based on the no-action-at-a-distance 
 assumption. 
 
 The 
 important point in establishing the EPR paradox for this more 
 general yet practical situation is that under the EPR premises 
  the predetermined statistics (or generalised ``elements of reality'') for the 
  physical quantities $\hat{x}$ and $\hat{p}$ 
 are attributed {\it simultaneously} to the subsystem at $A$. 
  Assuming no action-at-a-distance, the choice of the 
 experimenter (Bob) at $B$ to infer information about either $\hat{x}$ or 
 $\hat{p}$ 
 cannot actually induce the result of the measurement at $A$. As 
 there is no disturbance created by Bob's measurement, the
   (appropriately extended) EPR 
 definition of realism is that   
 the prediction for $x$ is something (a probabilistic ``element of reality'') that 
 can be attributed to 
 the subsystem at $A$, whether or not Bob makes his measurement. This 
 is also true of the prediction for $\hat{p}$, and therefore the two 
 ``elements of reality'' representing the physical 
 quantities $\hat{x}$ and $\hat{p}$ exist to describe 
 the predictions for $\hat{x}$ and $\hat{p}$  simultaneously. 
 
 The paradox can then be established by proving the 
 impossibility of such a simultaneous level of prediction for both $\hat{x}$ 
 and $\hat{p}$ for any quantum description of the subsystem $A$ alone. 
 By this we mean explicitly that there can be no procedure allowed, 
  within the predictions of quantum 
 mechanics, to make {\it simultaneous inferences} by measurements performed 
 at $B$ or any other location, of both the result $\hat{x}$ and 
 $\hat{p}$ at $A$, to the precision indicated by 
 $\Delta_{i}x<1$, $\Delta_{j}p<1$. (Recall that the inference of the 
 result at $A$ by measurement at $B$ is actually a measurement of 
 $\hat{x}$ performed with the accuracy determined by the $\Delta_{i}x$. 
 Simultaneous measurements of $\hat{x}$ and $\hat{p}$ to the accuracy 
 (\ref{eqn:eprcondeq}) are not possible (predicted by quantum 
 mechanics). The reduced density matrix 
 describing the state at $A$ after such measurements would violate 
 the H. U. P.)  

  More generally in an experiment we could obtain a 
mixed situation, where for example, 
the conditional variance $\Delta_{i}^{2}¥x$ for some of 
the $i$ might be greater than $1$. {\it The exact boundary at which we can 
claim an EPR paradox needs defining.}

   A simpler quantitative, experimentally testable criterion for EPR was proposed in 
 1989~\cite{eprr}.
  The 1989 inferred H.U.P. criterion~\cite{eprr} is based on the 
  average  variance of the conditional distributions 
 for inferring the result of measurement $\hat{x}$ (and also for 
 $\hat{p}$). The EPR paradox is demonstrated when 
 the product of the average errors in the 
 inferred results for 
 $\hat{x}$ and $\hat{p}$ violate the corresponding Heisenberg 
 Uncertainty Principle. The spirit of the 
 original EPR paradox is present, in that one can perform a measurement on $B$ 
 to enable an estimate of the result $x$ at $A$ (and 
 similarly for $\hat{p}$). 
 
  We define
\begin{eqnarray}
   \Delta_{inf}^{2}¥x&=&\sum_{i}P(x_{i}^{B}¥) \Delta 
   _{i}^{2}x\nonumber\\
   \Delta_{inf}^{2}p&=&\sum_{j}P(p_{j}^{B}¥) \Delta _{j}^{2}p
   \end{eqnarray}
  Here $\Delta_{inf}^{2}¥\hat{x}$ is the average variance for the prediction (inference) 
 of the result $x$ for $\hat{x}$ at $A$, conditional 
  on a measurement $\hat{x}^{B}$ at $B$. Here $i$ labels all outcomes 
  of the measurement $\hat{x}$ at $A$, and  $\mu_{i}¥$ and 
 $\Delta_{i}x$  are 
 the mean and standard deviation, respectively, of the conditional distribution 
 $P(x|x_i^{B}¥)$, where $x_{i}^{B}¥$ is the result of the measurement 
 $\hat{x}^{B}$ at $B$.
   We define a $\Delta_{inf}^{2}¥\hat{p}$ similarly
   to represent the weighted variance for the prediction (inference)
    of the result 
   $\hat{p}$ at $A$, based on the result of the measurement at $B$.  
   Here $P(x_{i}^{B})$ is the probability for a 
 result $x_{i}^{B}$ upon measurement of $\hat{x}^{B}$, and $P(p_{j})$ is 
 defined similarly.

 The criterion to demonstrate the EPR paradox, the
     signature of the EPR paradox, is  
   \begin{equation}
   \Delta_{inf}x
    \Delta_{inf}p \label{eqn:eprcritt}
  < 1.
  \end{equation}
 This 
    criterion is  a clear criterion for the demonstration of the EPR 
    paradox, by way of the argument presented above. Such a prediction 
    (\ref{eqn:eprcritt}) for $\hat{x}$ and $\hat{p}$ with 
    the average inference variances given, cannot be achieved by any 
    quantum description of the subsystem alone. This EPR criterion has  
    been achieved 
    experimentally. It is most useful in the proof of security for 
the quantum cryptographic proposal discussed in 
the companion paper~\cite{mdrcrytwo}.

\subsection{Best estimates criterion}

It is not always convenient to measure each conditional distribution 
$P(x/x_{i}^{B})$ and $P(p/p_{j}^{B})$. It is possible to construct 
other measurements based on the measurement of a sufficiently reduced noise 
in the appropriate sum or difference $\hat{x}-g\hat{x}^{B}¥$ and  
$\hat{p}+g\hat{p}^{B}$ (where here $g$ is a number). This sort of 
measurement was proposed in~\cite{eprr} in 1989 as a means to demonstrate EPR 
correlations, and is closely linked to squeezing and 
quantum-nondemolition measurements.

 We therefore consider a more general situation where an estimate or prediction $x_{est}$ is given of 
 the result for $\hat{x}$ at $A$, based on a result $x_{i}^{B}$ at 
 $B$ for the measurement $\hat{x}^{B}$. For each result $x_{i}^{B}$ 
 we define the rms error 
 \begin{equation}
  \delta_{i}^{2}=\sum_{x}¥P(x/x_{i}^{B})(x-x_{est})^{2}.
 \end{equation}
  The average error in the inference based on the 
  particular  
 estimate is given by
 \begin{eqnarray}
  \Delta_{inf,est}\hat{x}&=&\sum_{i}¥P(x_{i}^{B})\delta_{i}^{2}\nonumber\\
  &=&\sum_{x,i}P(x,x_{i}^{B})(x-x_{est})^{2}.
  \end{eqnarray}
 The  best estimate  
 (meaning that which will minimize the root mean square error  
 ~\cite{stat}) of the outcome of 
 $\hat{x}$ at $A$, based on a result $x_i^{B}¥$ for the   
 measurement at $B$, is given when $x_{est}=\mu_{i}$, since this 
 minimizes each $\delta_{i}^{2}$. The quantity 
 \begin{eqnarray}   
 \Delta_{inf}^{2}\hat{x}&=&\sum_{x,i}¥P(x_{i}^{B})\Delta_{i}^{2}\nonumber\\
 &=&\sum_{x,i}¥P(x,x_{i}^{B})(x-\mu_{i})^{2}
 \end{eqnarray}
  then defines the minimum average variance 
  for   
  the inference of the result of a 
 measurement $\hat{x}$ at $A$, based on the result of the measurement 
 $\hat{x^{B}¥}$ at $B$.
   Generally 
   \begin{equation}
   \Delta_{inf,est}\hat{x}\geq\Delta_{inf}\hat{x} \label{eqn:averagegreat}
   \end{equation}
    and also $\Delta_{inf,est}\hat{p} 
 \geq\Delta_{inf}\hat{p}$.
 
The observation of a violation of the ``inferred Heisenberg Uncertainty 
principle''
\begin{equation}  
\Delta_{inf,est}\hat{x}\Delta_{inf,est}\hat{p}<1 \label{eqn:eprcrit}
\end{equation} 
is a demonstration of the EPR paradox.
 Here  
 ``elements of
 reality'' ${\tilde x}$, ${\tilde p}$ 
simultaneously attributed to 
system $A$ by local realism would have a degree of definiteness not 
compatible, in the EPR sense, with the uncertainty principle. The criterion 
(\ref{eqn:eprcrit}) is a violation of the inferred Heisenberg 
Uncertainty Principle, 
a signature of an EPR paradox experiment defined in 1989.

\subsubsection{Linear estimation}

 As an example of the 1989 criterion, 
  we propose 
 upon a result $x_{i}^{B}¥$ for the measurement at $B$ that the predicted 
 value for the result $x$ at $A$ is given linearly by the estimate  
 $x_{est}=gx_{i}^{B}¥+d$. 
 For example 
 suppose our two systems at $A$ and $B$ are harmonic oscillators, with 
 boson operators $\hat{a}$ and $\hat{b}$ respectively, and 
 become correlated as a result of a coupling described by the 
 interaction Hamiltonian  
 $H_I = i\hbar \kappa(\hat{a}^{\dagger} \hat{b}^{\dagger} -\hat{a}\hat{b})$, 
  which acts for a finite time $t$.  For vacuum initial states this 
 interaction generates two-mode squeezed light~\cite{cavessch} 
\begin{eqnarray}
	|\psi> = \sum_{n=0}^{\infty}
	c_{n}¥|{n}>_{a} |{n}>_{b} \label{eqn:twomode}
	\end{eqnarray}
 where $c_{n}= tanh^{n}r/cosh r$ and $r=\kappa t$. This simple quantum 
 state was shown to be EPR-correlated in reference ~\cite{eprr}, 
and has been
 used to  model  continuous-variable EPR-correlated fields  
 to date.
 Here we define the quadrature phase amplitudes 
 \begin{eqnarray}
		\hat{x}&=&\hat{X}_a =(\hat{a}+\hat{a}^\dagger)\nonumber\\
		\hat{p}&=&\hat{P}_a= (\hat{a}-\hat{a}^\dagger)/i\nonumber\\ 
		\hat{x}^{B}&=&\hat{X}_b =(\hat{b}+\hat{b}^\dagger)\nonumber\\
		\hat{p}^{B}&=&\hat{P}_b= 
		(\hat{b}-\hat{b}^\dagger)/i.\label{eqn:quad}
		\end{eqnarray}
The Heisenberg 
uncertainty relation for the orthogonal 
amplitudes of mode $\hat{a}$ is 
 $\Delta^{2}X_{a}¥\Delta^{2}P_{a}¥\geq  1$.

 The size of the 
deviation $\delta_{i}¥=x-(gx_{i}^{B}¥+d)$ in the linear estimate 
$x_{est}$ can then
 be measured (or 
calculated). We simultaneously measure $\hat{x}$ at $A$ and 
$\hat{x}^{B}¥$ at $B$, to determine $x$ and $x_{i}^{B}¥$ and then to 
calculate initially for given $x_{i}^{B}¥$ 
$\langle\delta^{2}\rangle_{i}¥=\sum_{x}¥ 
P(x,x_{i}^{B}¥)(x-(gx_{i}^{B}¥+d)^{2})/
 P(x_{i}^{B}¥) $.  Averaging over the different values of $x_{i}^{B}¥$ we obtain as a measure 
of error in our inference, based on the linear 
estimate: 
\begin{eqnarray}
\Delta_{inf,L}^{2}\hat{x}&=&\sum_{x_{i}^{B}¥}P(x_{i}^{B}¥)
\langle\delta^{2}\rangle_{i}\nonumber\\
&=&
\sum_{x,x_{i}^{B}¥}P(x,x_{i}^{B}¥)\{x-(gx_{i}^{B}¥+d)\}^{2}\nonumber\\
&=&\langle\{\hat{x}-(g\hat{x}^{B}¥+d)\}^{2}\rangle.
\end{eqnarray}
 The best linear estimate $x_{est}$ is the one that will minimize  
$\Delta^{2}_{inf,L}\hat{x}$.  This corresponds to the choice~\cite{stat} 
$d=-\langle (\hat{x}-g\hat{x}^{B}¥)\rangle$. Denoting 
$\delta_{0}=\hat{x}-g\hat{x}^{B}¥$, our choice of estimate optimized with 
respect to $d$ gives a minimum error 
\begin{equation}
\Delta^{2}_{inf,L}\hat{x}=
\langle \delta^{2}_{0}¥-\langle\delta_{0}¥\rangle^{2}¥\rangle.
\end{equation} 
The best choice for $g$ is discussed in~\cite{eprr}. 
 The two-mode squeezed state will 
 predict 
~\cite{eprr} ($g=\tanh 2\kappa t$) the correlations 
$X_{a}=X_{b}$, and $P_{a}=-P_{b}$ to give 
\begin{eqnarray}
	\Delta_{inf,L}^{2}\hat{x}=\Delta_{inf,L}^{2}\hat{p}=1/\cosh 
	2\kappa t
\end{eqnarray}
The EPR correlations are predicted possible for all nonzero values of the 
two-mode squeeze parameter $r$.

If the estimate $x_{est}¥$ corresponds to the mean of the 
conditional distribution 
$P(x|x_{i}^{B}¥)$ then 
the variance $\Delta^{2}_{inf,L}\hat{x}$ will correspond to 
the average conditional variance $\Delta_{inf}^{2}x=\sum_{x_{i}^{B}¥}P(x_{i}^{B}¥)\Delta_{i}¥^{2}¥$ 
specified in Section 3b above. This is the case, with a certain choice of $g$, for the two-mode 
squeezed state used to model continuous variable 
EPR states generated to date.

In general the 
variances of type $\Delta^{2}_{inf,L}\hat{x}$ 
based on estimates will be greater than or equal to the optimal
 evaluated from the conditionals (this was shown  
  in Section 3b):  we have $ \Delta_{inf,L}\hat{x}\geq 
  \Delta_{inf}\hat{x}$ and $ \Delta_{inf,L}\hat{p}\geq 
  \Delta_{inf}\hat{p}$.
   The observation of
   \begin{equation}  
    \Delta_{inf,L}\hat{x}\Delta_{inf,L}\hat{p}< 1
    \end{equation}
      by way of the 1989 EPR criterion 
        (\ref{eqn:eprcrit}), will then also 
        imply the situation of the EPR paradox.

  \subsection{Generalised hidden variable EPR criteria}

   Our objective was to determine the boundaries at which one could 
   sensibly claim EPR correlations. 
   To determine a whole set of more general EPR criteria, we first 
   need to consider  what is 
   meant by the assumption of ``local realism''
   as applied to  less strongly correlated 
   subsystems, and to then define situations where this assumption 
   would imply an ``incompleteness'' of quantum mechanics. EPR's 
   ``realism'' was defined originally only in the context of perfect 
   correlation, and therefore it must be pointed out that 
   any generalization actually goes further 
   than EPR. 
     
 We propose then that the {\it most  
  general application of the EPR premise of local realism to  
  correlated systems is to simply postulate the existence of a some 
   local realistic 
  theory  to describe the correlations}. If we can show that in order to 
  predict the quantum correlations correctly, we {\it need} to use local realistic 
  states (``elements of reality'') that  {\it predetermine with sufficient  definiteness the results 
  of $\hat{x}$ and $\hat{p}$ measurements}, so that the level of 
  definiteness cannot be represented by a quantum local state for the 
  subsystem,  then 
  we have the situation of the EPR argument in its most general sense. 
  Namely, that if the premise of local realism is valid, then we need 
  to ``complete'' quantum mechanics (to introduce hidden variables) to 
  correctly predict the quantum statistics.

    EPR's ``realism''  implied, for the perfect predictions 
   they considered, the 
   existence of 
   variables that predetermine with an 
   absolute definiteness the predictions for the results of 
   measurements $\hat{x}$ and $\hat{p}$. These variables are `hidden 
   variables'' since they cannot be represented by any quantum state, 
   and therefore exist to supplement (complete) quantum mechanics. 
   While EPR themselves did not use the 
   term ``hidden variable'', it is generally understood that this is 
   the consequence of their argument~\cite{CH}, the hidden variable 
   being the mathematical symbol of the ``element of reality''.

  EPR's argument is based on the assumption of the validity of a local 
  realism defined only with respect to perfectly correlated systems. We 
  propose the following, that {\it the most general 
  meaning of the assumption of 
  local realism, as applied to any more weakly correlated 
  system, is the ability to correctly 
  describe the predictions of measurements through some actual local realistic 
  theory, and that such theories are symbolised mathematically 
  by a general ``local hidden 
  variable theory'' of the type considered by Bell~\cite{Bell,CH}}.
  
  The EPR criteria derived from this proposal are quite 
  different to the constraints (Bell inequalities) derived by Bell, 
  which are derived from the assumption that a local hidden variable theory  
  will correctly predict the experimental outcomes. The distinction 
  between EPR and Bell tests is discussed in 
  Section 4.

  However {\it it must be stressed} that it was EPR's intention (presumably) 
  to discuss the 
  incompatibility of the  
  ``completeness of quantum mechanics'' with ``local realism'' (to 
  imply the need for hidden variables) in an 
  intuitive, physical way, {\it  without introducing any particular 
  mathematical formulation for the 
  meaning of local realism}. This is an {\it important philosophical and 
  historical point}. 
  In this regard the criteria discussed 
  in Sections 3a,b, and c above must be regarded as stronger EPR criteria, 
  necessary for the demonstration 
  of the EPR paradox itself. It is mentioned also at this point that 
  the previous EPR criteria are not only stronger in this 
  philosophical sense, 
  but are mathematically stronger.  By this we mean that 
  the EPR criteria given in Sections 3a,b and c are a subset of the more 
  general EPR criteria we derive here, as will be shown in Section 
  5~\cite{weak}. 
  
  To show however that the criteria we consider in this section still fit 
  into the category of general EPR criteria, it is simply argued 
  that if  local realism as applied 
  to systems of arbitrary correlation (the starting 
  point of a generalised EPR argument) is to be valid, then there is a 
  need to propose a
   working mathematical theory satisfying local realism in order to 
  predict the results of experimental measurements.  
   The question then is whether any kind of local realism can be 
   constructed using something different to the mathematical basis of 
   the local realistic theories studied by Bell. To my knowledge there 
   is no such mathematical representation~\cite{weak}.
   
  In order to derive mathematical criteria, we formulate the 
  mathematical description of local realism along the lines of 
  Bell~\cite{Bell,CH}. Such general local realistic theories do not ``a 
  priori'' put any restrictions~\cite{fuzzy} on the degree of definiteness of the 
  prediction for the results of measurements, for a system described 
  by a given local realistic 
  (hidden variable) state. 
  For  more weakly correlated systems, 
  since the result at 
   $A$ is not completely determined by the result at $B$,
   we introduce local realistic (local hidden?) variables $\{\lambda\}$ 
   to symbolize possible states of the subsystem $A$, 
   but where now {\it for 
   each such local variable specification of the subsystem, 
   there can 
    be   
   an  unspecified fuzziness, symbolized by the variances 
   $\Delta_{\lambda}^{2}¥x$ and 
 $\Delta_{\lambda}^{2}¥p$ that give 
   the predicted values for 
   the result of the $\hat{x}$ or $\hat{p}$ measurement 
   respectively}~\cite{disc}. 
   It is interesting that the concept of a ``blurred'' reality 
   was discussed qualitatively by 
   Schrodinger~\cite{sch} in his reply to EPR in 1935. This approach 
   is illustrated in Figure 1.

 At $A$ there is the choice to measure either $\hat{x}$ 
or $\hat{p}$, a choice denoted by different values, $0$ and $\pi/2$ 
respectively, of the 
parameter $\theta$. Similarly at $B$ there is the choice, denoted by $\phi$,
 to measure $\hat{x}^{B}¥$ or 
$\hat{p}^{B}¥$.  The 
 (hidden) variable values $\{\lambda\}$
  determine the results, or probabilities for 
results, of measurements if performed. 
There will be a probability $p_{x}^A(\theta, \lambda )$ 
for the result $x$ of a 
measurement $\theta$ at $A$, given the hidden variable state 
$\{\lambda\}$; similarly a $p_{y}^B(\phi, \lambda )$ is defined. 
 The
$\Delta^{2}_{\lambda}x$, $\Delta^{2}_{\lambda}p$ are the variances of
 $p_{x}^{A}(\theta=0,\lambda)$, $p_{x}^{A}¥(\theta=\pi/2,\lambda)$ 
 respectively. In 
accordance with EPR's locality (no action-at-a-distance) assumption, this  
probability  distribution is independent of the 
 experimenter's choice $\phi$ of simultaneous 
 measurement at $B$. 
  Assuming a general local 
hidden variable theory the joint probability 
$P_{\theta,\phi}(x,x^{B}¥)$ of obtaining an outcome $x$ 
at $A$ and $x^{B}¥$ at $B$ is of the form ($\rho(\lambda)$ is the 
probability distribution for the $\{\lambda\}$) 
\begin{eqnarray}
P_{\theta ,\phi}(x,x^{B}¥)= \int_{\lambda}¥ \rho(\lambda) \quad 
p_{x}^A(\theta, \lambda ) 
p_{x^{B}¥}^B(\phi, \lambda )\quad d\lambda  \label{eqn:bellsep}
\end{eqnarray}
  These general local hidden variable theories were 
  considered by Bell~\cite{Bell}.

 \begin{figure}
    \includegraphics[scale=.6]{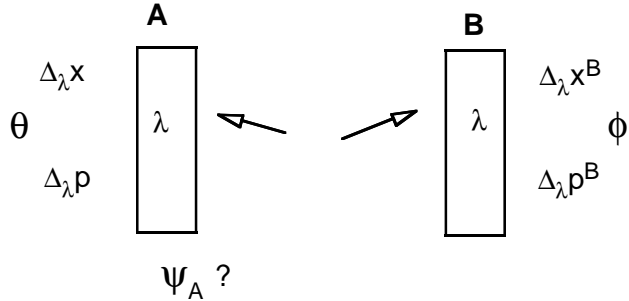}
\caption[fig1]
{The local realist will describe the correlated statistics through 
classical variables $\lambda$ (that can be shared by the two 
subsystems at $A$ and $B$). These classical variables depict possible 
states of the subsystems. The  $\Delta_{\lambda}^{2}¥x$ and 
$\Delta_{\lambda}^{2}¥p$ are the variances of the distributions 
$p_{x}^{A}(\theta=0,\lambda)$ and $p_{x}^{A}(\theta=\pi/2,\lambda)$, 
respectively, that 
denote the probability of a  result $x$ upon measurement of  
$\hat{x}$  ($\theta=0$) or 
$\hat{p}$ ($\theta=\pi/2$), given that the system is in a 
particular state $\lambda$. Similar definitions 
exist  for the subsystem at $B$. The system is EPR-correlated in a 
general sense when it 
can be demonstrated that in order to predict the observed statistics, 
 $\Delta_{\lambda}x  
\Delta_{\lambda}p<C$ (where $\Delta \hat{x} \Delta \hat{p}\geq 1 $ 
is the uncertainty relation) for at least one of the possible   
states $\lambda$. }% 
\label{fig1.eps}
\end{figure}

    In  order to  
    demonstrate EPR correlations, or the EPR paradox in the 
   most general way, we assume the validity of a local realistic 
   description and then must {\it prove the necessity 
   of} using hidden variable states with a simultaneously well-defined 
   prediction for ``position'' and ``momentum'', so that  
   $\Delta_{\lambda}x\Delta_{\lambda}p
    <  1$.
   To test for EPR correlations, the objective then 
   is {\it to demonstrate the 
  failure of any local realistic  theory (\ref{eqn:bellsep})  
  satisfying the auxiliary assumption
   \begin{equation} 
 \Delta_{\lambda}x\Delta_{\lambda}p
    \geq  1\label{eqn:prov}
    \end{equation}  
 (and a similar restriction is placed on the variances 
   of the $p_{x^{B}¥}^B(\phi, \lambda )$) to describe the statistics of the correlated, spatially-separated 
    fields}. 
     This would indicate that at least part of the time, if local 
     realism, as manifested through any actual local realistic theory, 
 is to correctly describe the statistics, 
  the local subsystem at $A$ must be in a truly hidden variable state satisfying 
    $\Delta_{\lambda}x\Delta_{\lambda}p
    <  1$, a description 
 that cannot be given by a local  
 quantum wavefunction for the subsystem alone. 
 The ``incompleteness of 
 quantum mechanics'' from the point of view of the local realist
  then follows: he/she must introduce ``hidden'' variables to 
  ``complete'' quantum mechanics so as to enable a local realistic 
  description.    
  This is the situation of 
  EPR correlations in their most generalized (weakest) sense.

  To derive criteria sufficient to demonstrate EPR correlations, we 
  assume the general local realistic description 
 (\ref{eqn:bellsep}) 
  with the proviso (\ref{eqn:prov}) and consider the whole set of 
  inequalities or other constraints following necessarily 
  from these assumptions.  Such constraints are derived explicitly in Section 
  5, where it is also shown that these are constraints sufficient to 
  quantum inseparability.

  \subsection{Summary of EPR criteria}
  
  We classify two types of EPR correlations. The first 
  type are based on variances of conditional distributions reflecting 
  the ability to infer a result for measurement at $A$ based on a 
  measurements at spatially separated location $B$. These strong EPR 
  correlations are evidenced through EPR criteria such as that of 
  1989, the inferred H. U. P. EPR criterion  
  (\ref{eqn:eprcrit}), and are 
   necessary for the demonstration of EPR correlations defined 
   in the spirit of  
   the original EPR paradox (where the premise of local realism is defined and 
   employed with no  
  assumptions made regarding the mathematical form a local realistic 
  theory would take). 
  In terms of quadrature phase amplitude measurements 
  this strong EPR criterion is satisfied when
     \begin{equation}
      \Delta(X_{a}-gX_{b})\Delta(P_{a}+hP_{b})<1\label{eqn:strong}
      \end{equation}
   where the quadrature amplitudes have been defined in Section 3c, 
   and $g,h$ are parameters chosen to minimize the variances.

  The second generalised and weaker type~\cite{weak} of EPR correlations are demonstrated through the 
  failure of local realistic (Bell-type local hidden variable) descriptions 
  to predict the measured statistics of the spatially 
  separated fields, unless we allow for a degree of 
  definiteness in their  prediction for results of an $\hat{x}$ or 
  $\hat{p}$ 
  measurement. These correlations demonstrate that if we do assume the 
  correctness of local realism through the existence of an actual 
  (Bell-type) local 
  realistic theory, then the hidden variables describing each local 
  subsystem {\it must} give such definite predictions for position and 
  momentum that they cannot also be represented by a local quantum state for 
  that subsystem.  
In  this way an inconsistency between the assumption 
  of local realistic theories and the completeness of quantum 
  mechanics is established. Although not based on conditional 
  measurements, these weaker correlations 
  therefore  still reflect the essence of the EPR argument. 
  
  The 
 explicit forms for some of these weaker constraints are derived in Section 5.
  It is also shown in Section 5 that this set of constraints 
 includes all of the stronger 1989 inferred H. U. P. 
 EPR criteria discussed above. An example of a weaker EPR criterion, 
 which is not a stronger EPR criterion, is derived in Section 5c and is written 
 in terms of  
 quadrature phase amplitude measurements as
      \begin{equation}
      \Delta(X_{a}-X_{b})\Delta(P_{a}+P_{b})<2
      \end{equation}
This criterion is closely related to the criterion for the 
observation of a two-mode squeezing. 
Two-mode squeezing is said to be observed when  
     $\Delta^{2}¥(X_{a}-X_{b})<2$, or 
 $\Delta^{2}¥(P_{a}+P_{b})^{2}<2$. In the limit of perfect 
 correlation, the optimal $g,h$ values for (\ref{eqn:strong}) are 
 $1$, and here the weaker nature of the last criterion is apparent. 
 The constraints sufficient to prove quantum inseparability derived 
 recently by 
 Duan et al~\cite{content} and Simon~\cite{content} 
 are also examples of generalised (weaker) EPR criteria.

 \section{Significance of the demonstration of the EPR 
 correlations and relationship to Bell inequalities}

In order to appreciate the significance of an experimental 
 demonstration of EPR, 
 it is crucial  to realize that local realistic 
 descriptions may be entirely compatible, depending on the 
 statistics, with a local quantum 
 description for each of the subsystems $A$ and 
 $B$. Where EPR correlations cannot be proven, the observed 
statistics could be explained using   
local variables or parameters to represent possible states of the 
subsystem that have  an intrinsic fuzziness associated with 
their predictions for measurements, so that this fuzziness could be entirely 
 the result of a quantum state for each localized subsystem.

A local realist takes the viewpoint that (hidden) 
variables exist to predetermine (but perhaps with some unspecified 
indeterminacy) 
the result of measurements for 
subsystem $A$ in a way which is compatible with EPR's no 
action-at-a-distance (locality).  But at the 
onset of the experimental proof of EPR correlations, the local 
realist's  view is one 
that ``completes'' quantum mechanics, since the (hidden) variables 
representing the local subsystems  define the result for measurement
 so precisely that 
they cannot also represented by a quantum state for the subsystem
 $A$ or $B$, itself.  In this manner, {\it at the onset of EPR 
 correlations, a dispute is established between 
the local realist and the believer in the ultimate completeness of 
the quantum state}; though not between the local realist and the 
believer in quantum  mechanics (presumably EPR were in this category) 
who is perfectly prepared to supplement quantum mechanics
with hidden variables.

  From the local hidden variable form (\ref{eqn:bellsep})  
  various constraints (called Bell-type inequalities~\cite{Bell}) may be 
  derived, that do not require the assumption of any auxiliary 
  assumptions such as (\ref{eqn:prov}). 
 Where one shows a violation of such a constraint (which is also an 
 example of a  
 separability criterion), then we have a violation of a Bell-type inequality. 
 From such a violation one may draw the stronger conclusion than can 
 be drawn from a demonstration of EPR paradox itself. In the case 
 where one violates a Bell inequality, it is proved 
 that local realism (or all local hidden variable theories) 
 itself is invalid,  and that the predictions of the particular quantum states 
 in this case cannot be equivalently represented by any more ``complete'' 
 theory (in which hidden variables are introduced) still 
 satisfying local realism.

The two-mode squeezed state (\ref{eqn:twomode}) 
 predicts EPR correlations~\cite{eprr} satisfying the EPR condition 
 $\Delta_{inf}\hat{x} 
  \Delta_{inf}\hat{x}< 1$. However it is well-known that 
  the joint probabilities $P_{\theta ,\phi}(x,x_{i}^{B}¥)$
 for the results of measurements 
$\hat{x},\hat{p}$ 
can be predicted from a local hidden variable theory  
(\ref{eqn:bellsep}), derived from the  Wigner function, which is positive for the 
two-mode squeezed state. In such a local hidden theory the Wigner function c-numbers $x$ and $p$ 
take on the role of position and momentum hidden variables;   
the Wigner function, being positive, gives the 
probability distribution for the hidden variables $x,p$. 
This implies that the  Bell inequalities will not be violated in this 
case for direct $\hat{x},\hat{p}$ 
measurements. (Of course Bell inequalities may be violated for other 
measurements (not $\hat{x}$, $\hat{p}$) on this quantum state; this 
is a different issue, however).

The existence of a local hidden variable theory that would predict 
 the EPR $\hat{x}$, $\hat{p}$ correlations  
 might lead to the interpretation that the EPR 
 experiment, demonstrating $\Delta_{inf}\hat{x} 
  \Delta_{inf}\hat{x}< 1$,  
reflected a situation in which quantum and local realistic (classical)
 domains are not 
distinguishable, and that ``there is no real paradox''. 
This is not the case. 

 The local realistic 
hidden variable theory 
used to give the quantum predictions    
is, necessarily, not actually quantum theory, since it  must 
incorporate a description $\{\lambda_{a}\}$ 
for a state of the subsystem at $A$ or $B$ 
in which the $x$ and $p$ are prespecified to a variance better than 
the uncertainty principle. 
At the proof of demonstration of EPR correlations,  the 
quantum mechanics theorist  {\it cannot generate a 
separable description} in which each subsystem can be represented 
locally by a 
quantum state; his/her quantum state is 
necessarily entangled. (This is shown in Section 5). 
The separable local hidden variable theory 
based on the Wigner function is not a separable (local) theory in 
quantum mechanics since these simultaneously well-defined $x$ and 
$p$ are not quantum states. 

The philosophical 
viewpoint of the local realist can  be compatible with the 
quantum description {\it only if} quantum mechanics is ``completed'' 
 to incorporate the local hidden variables ( for example, so that the 
 c-numbers of the quantum Wigner function are the positions and momenta 
 of the particle) and to therefore 
 provide a  
local (separable) description. The {\it conflict between the local 
realist and the believer in the completeness of quantum mechanics 
still exists, and there is  the paradox of EPR, in spite of the fact 
that there is a local realistic description for the quantum 
predictions in this case}.

 \section{Link with entanglement criteria }
 
   The EPR criteria are closely linked to criteria for entanglement. 
   The demonstration of entanglement may be defined as the measured 
  experimental violation of any one of the  
  set of criteria following necessarily from the assumption of
   separability, where  
  a separable quantum state is   
 defined as being expressible by a 
 density matrix of the form
     \begin{equation}
     \rho= \sum_{\lambda}P_{\lambda}\rho_{\lambda}^{A}\rho_{\lambda}^{B} \label{eqn:sep}
\end{equation}
 where $\sum_{\lambda} P_{\lambda}=1$. Here $\lambda$ is simply a discrete or 
 continuous label for quantum states, and no longer refers to hidden variables.
 Necessary conditions for separability for finite-dimensional systems 
 were explored by Peres and 
 Horodecki et al~\cite{entsep}.
  Such necessary conditions using continuous variable measurements
   have been derived 
      by Duan et al~\cite{content} and 
      Simon~\cite{content}. Here we are concerned with criteria 
      expressed, as are Bell inequalities, in terms of measurable 
      expectation values or probabilities for results of experimental 
      measurements that may be performed on the system.

   The prediction $P_{\theta,\phi}(x,y)$ given a separable quantum state  
 is of the form 
\begin{eqnarray}
P_{\theta ,\phi}(x,y)= \sum_{\lambda}¥ P_{\lambda}¥ \quad 
p_{x}^A(\theta, \lambda ) 
p_{y}^B(\phi, \lambda )  
\end{eqnarray}
where $p_{x}^A(\theta, \lambda )=\langle x|\rho_{\lambda}^{A}|x\rangle$ and 
$p_{y}^B(\phi, \lambda )=\langle y|\rho|y\rangle$ and here $|x\rangle$ and 
$|y\rangle$ are eigenstates of the operators representing 
measurements at $A$ and $B$ respectively. 
 In the Sections 5b and c below, we prove certain  constraints to follow 
  necessarily from the assumption of quantum separability (\ref{eqn:sep}). 
  The violation 
  of these constraints is then sufficient to demonstrate entanglement.

The predictions 
based on quantum separability must allow for all possible quantum 
density 
operators and therefore we  
make no further assumptions except to assume the separable nature of the 
decomposition, and the restriction put on the statistics due to 
general quantum bounds such as the uncertainty relation satisfied by 
each local quantum state $\rho_{\lambda}^{A}$ and $\rho_{\lambda}^{B}$.

 In the proofs given in Section 5b and c,
  we assume the separable form (\ref{eqn:sep}) and apply the 
  constraint that the local quantum density operator must allow the 
  Heisenberg 
  uncertainty principle, to derive the necessary separability 
  criteria.   Using the same algebra, {\it the same constraints are 
  derivable from the local hidden variable form (\ref{eqn:bellsep})     
 with the auxiliary 
  constraint (\ref{eqn:prov}), meaning that violation of 
  these constraints will also 
  imply EPR correlations in the most general sense}, as defined in 
  Section 3d.
In this 
way it is seen that {\it the generalised EPR constraints are identical to 
those derived as necessary conditions of separability where the 
additional quantum restriction in connection with the uncertainty 
relation is assumed}.

  Other separability criteria, using the uncertainty product bound, 
have been derived by Duan et al and Simon~\cite{content}. These criteria are 
important because they have been shown to be necessary and 
 and sufficient to demonstrate entanglement for certain quantum states 
 (Gaussian states) relevant to many experimental situations. 
 These constraints also follow from the assumptions (\ref{eqn:bellsep})     
  and (\ref{eqn:prov}), and are thus examples of generalized  EPR 
  criteria, as defined in Section 3d.
 
   In deriving a particular constraint from the assumption of quantum 
   separability (\ref{eqn:sep}), one {\it either} derives a result based 
   on (\ref{eqn:sep}) (or (\ref{eqn:bellsep})) alone, {\it or else} makes 
   further assumptions 
   regarding a general property of a quantum state $\rho_{\lambda}^{A}$ such as 
   satisfaction of the uncertainty relation (or 
   (\ref{eqn:prov})). The 
   constraints derived based only on  (\ref{eqn:sep}) (or (\ref{eqn:bellsep}) 
   are called Bell-type inequalities and their violation is therefore 
   proof of  failure of all local hidden variable theories 
   (\ref{eqn:bellsep}), as well as being demonstrations of 
   inseparability. The violation of the remaining constraints (that are 
    based on an additional assumption), cannot imply failure of all 
   local hidden variables, but nonetheless may be classified as a 
   violation of a generalised 
    EPR criterion.

   To summarize, {\it provided measurements and  
   spatial separations between subsystems $A$ and $B$ allow   
  justification of the locality assumption}, 
   the measured violations of one of the necessary criteria for separability 
   (where an extra constraint relating to general quantum  bound such 
   as the uncertainty relation is 
   assumed) are not only a demonstration of entanglement, but  
  are then none other than a demonstration of 
  EPR correlations in the generalized sense.

\subsection{EPR-criteria as signatures of entanglement}

If we demonstrate the EPR paradox in its  
generalized form, then it follows through the very meaning of the EPR 
paradox that we must use an entangled source and that demonstration of 
EPR correlations must imply entanglement. A
separable source as given by (\ref{eqn:sep}) has the interpretation 
that it is always in one of the 
 factorizable 
states $\rho_{\lambda}^{A}\rho_{\lambda}^{B}$ with probability $P_{\lambda}$; in each 
case the subsystem at $A$ being describable by the quantum state 
$\rho_{\lambda}^{A}$ and the subsystem $B$ being described by the 
quantum state $\rho_{\lambda}^{B}$. These states represent local 
descriptions where for each such description the predictions
(``elements of reality'') for the position and momentum measurements 
 are 
sufficiently indefinite so that the uncertainty bound 
(\ref{eqn:prov}) 
is satisfied. The predicted statistics must be compatible with this 
local fuzzy description, and in being so cannot satisfy the general 
EPR criterion discussed in relation to equation (\ref{eqn:prov}) 
in Section 3d.

\subsection{1989 Inferred H. U. P. EPR criterion as a signature of 
entanglement}

  We will 
 first 
prove that separability will always imply 
 $\Delta_{inf}\hat{x}\Delta_{inf}\hat{p}\geq 1$, meaning that a 
 satisfaction of the 1989 
 criterion as given by (\ref{eqn:eprcrit}) for EPR correlations will 
 always imply entanglement. (The same algebra is proof that the 
  strong 1989 EPR 
 criterion follows necessarily from the assumptions (\ref{eqn:bellsep}) and 
 (\ref{eqn:prov}), and therefore that the 1989 EPR criterion is also a 
 weaker EPR criteria of the type discussed in Section 3d.)

 The  
conditional probability of 
result $x$ for measurement $\hat{x}$ at $A$ given a simultaneous 
measurement of $\hat{x}^{B}$ at $B$ with result $x_i^{B}¥$ is   
$P(x|x_{i}^{B}¥)=P(x,x_{i}^{B}¥)/P(x_{i}^{B}¥)$ where, 
assuming separability (\ref{eqn:sep}), 
     \begin{eqnarray}
     P(x,x_{i}^{B}¥)
    =\sum_{\lambda}¥P_{\lambda}¥P_{\lambda}(x_{i}^{B}¥)P_{\lambda}(x) \label{eqn:begsep}
 \end{eqnarray}
Here $|x\rangle,|x^{B}¥\rangle$ are the eigenstates of 
$\hat{x}$,$\hat{x}^{B}¥$ 
respectively, and $P_{\lambda}(x)=\langle x|\rho_{\lambda}^{A}|x\rangle$, 
$P_{\lambda}¥(x_{i}^{B}¥)=\langle x_{i}^{B}¥|\rho_{r}^{B}|x_{i}^{B}¥\rangle$.
The mean $\mu_{i}¥$ of this conditional distribution is
\begin{eqnarray}  
     \mu_{i}&=&\sum_{x}xP(x|x_{i}^{B}¥)\nonumber\\
     &=&\{\sum_{\lambda}P_{\lambda}P_{\lambda}¥(x_{i}^{B}¥) 
     \langle x\rangle_{\lambda}\}/P(x_{i}^{B}¥)
\end{eqnarray}
 where $\langle x\rangle_{\lambda}=\sum_{x}xP_{\lambda}¥(x)$.
  The variance $\Delta_{i}^{2}x$ of the distribution $P(x|x_{i}^{B}¥)$ is 
\begin{eqnarray}
\Delta_{i}^{2}x
       &=&\{\sum_{\lambda}¥P_{\lambda}¥P_{\lambda}(x_{i}^{B}¥)
       \sum_{x}P_{\lambda}¥(x)
      (x-\mu_{i}¥)^{2}\}/P(x_{i}^{B}¥).
      \end{eqnarray} 
      For each state $\lambda$, the mean square 
deviation $\sum_{x}P_{\lambda}(x)(x-d)^{2}$ 
is minimized with the choice 
$d=\langle x\rangle_{\lambda}$~{\cite{stat}}.  
Therefore for the choice $d=\mu_{i}¥$,
\begin{eqnarray}
     \Delta_{i}^{2}x&\geq&
    \{\sum_{\lambda}¥P_{\lambda}¥P_{\lambda}(x_{i}^{B}¥)\sum_{x}
     P_{\lambda}¥(x)(x-\langle x\rangle_{\lambda})^{2}\}/P(x_{i}^{B}¥)\nonumber\\
     &=&\{\sum_{\lambda}¥P_{\lambda}¥P_{\lambda}¥(x_{i}¥)
     \sigma_{\lambda}^{2}¥(x)\}/P(x_{i}^{B}¥)
     \label{eqn:deltaconstraint}
\end{eqnarray}
 where $\sigma_{\lambda}^{2}¥(x)$ is the variance of $P_{\lambda}(x)$.
 Taking the average variance over the 
$x_{i}^{B}¥$ we get (recalling that for any set of statistics 
(\ref{eqn:averagegreat}) holds)  
\begin{eqnarray}
\Delta_{inf,est}^{2}¥\hat{x}&\geq&\Delta_{inf}^{2}\hat{x}\nonumber\\
&\geq&\sum_{x_{i}^{B}¥}P(x_{i}^{B}¥)\{\sum_{\lambda}P_{\lambda}¥
P_{\lambda}(x_{i}^{B}¥)
                \sigma_{\lambda}^{2}(x)\}/P(x_{i}^{B}¥)\nonumber\\
&=&\sum_{\lambda}P_{\lambda}\sigma_{\lambda}^{2}¥(x)
\sum_{x_{i}^{B}¥}P_{\lambda}(x_{i}^{B}¥)\nonumber\\
&=&\sum_{\lambda}P_{\lambda}\sigma_{\lambda}^{2}¥(x)
\end{eqnarray}
 Also $\Delta_{inf}^{2}\hat{p}\geq\sum_{\lambda}
  P_{\lambda}\sigma_{\lambda}^{2}¥(p)$, where 
 $\sigma_{\lambda}^{2}(p)$ is the variance of $P_{\lambda}(p)=\langle 
 p|\rho_{\lambda}^{A}|p\rangle$, $|p\rangle$ being the eigenstate of 
 $\hat{p}$. This implies (from the Cauchy-Schwarz inequality)  
  \begin{eqnarray}
  \Delta_{inf}^{2}\hat{x}\Delta_{inf}^{2}\hat{p}&\geq& 
  \{\sum_{\lambda}P_{\lambda}\sigma_{\lambda}^{2}¥(x)\}  
  \{\sum_{\lambda}P_{r}\sigma_{\lambda}^{2}¥(p)\}\nonumber\\
  &\geq&| \sum_{\lambda}P_{r}\sigma_{\lambda}¥(x)
   \sigma_{\lambda}(p) |^{2}.
   \end{eqnarray} 
 For any $\rho_{\lambda}^{A}$ it is constrained, by the uncertainty relation, that  
$ \sigma_{\lambda}(x)\sigma_{\lambda}(p)\geq 1$.
 We therefore conclude that for a 
separable quantum state
 \begin{equation}
  \Delta_{inf,est}\hat{x}\Delta_{inf,est}\hat{p}\geq
  \Delta_{inf}\hat{x}\Delta_{inf}\hat{p}\geq 1. \label{eqn:endsep}
\end{equation}
The experimental observation then of the EPR criterion 
 $\Delta_{inf}\hat{x}\Delta_{inf}\hat{p}< 1$ (or
 $\Delta_{inf,est}\hat{x}\Delta_{inf,est}\hat{p}< 1$), 
 as given by (\ref{eqn:eprcrit}),
will imply inseparability (that is entanglement).

 \subsubsection{Linear estimates}
 
In general the 
variances of type $\Delta^{2}_{inf,L}\hat{x}$ defined in Section 3c
based on linear estimates will be greater than or equal to the optimal
 (minimal) variances evaluated from the conditionals (this was shown  
  in Sections 3b and c):  we have $ \Delta_{inf,L}\hat{x}\geq 
  \Delta_{inf}\hat{x}$ and $ \Delta_{inf,L}\hat{p}\geq 
  \Delta_{inf}\hat{p}$.
   The separable state must then always 
 predict    $ \Delta_{inf,L}\hat{x}\Delta_{inf,L}\hat{p}\geq 1$
   and the observation of
   \begin{equation}  
    \Delta_{inf,L}\hat{x}\Delta_{inf,L}\hat{p}< 1
    \end{equation}
      implies
    quantum  inseparability, 
        for any $g$ and $d$.

      Given that the linear inference method of the previous 
      section is that so far  
      actually used in the two-mode 
      squeezing EPR experiments, it is worthwhile to demonstrate that 
   $\Delta_{inf,L}\hat{x}\Delta_{inf,L}\hat{p}< 1$ implies 
   inseparability explicitly. Such a proof is useful in that we 
   will also 
  prove that the observation of  two-mode 
   squeezing with respect to certain observables is
    sufficient to demonstrate entanglement (or a 
   generalized EPR correlation). Our proof, first presented in 
   ~\cite{qpr}, is from first principles, 
   although it has also been previously discussed by Braunstein et 
   al~\cite{brau} how the observation of 
    $\Delta_{inf,L}\hat{x}\Delta_{inf,L}\hat{p}< 1$ may imply 
    inseparability, based on criterion for inseparability derived by 
    Duan et al.

    Separability will imply, upon optimizing $d$ but 
   keeping $g$ general (use ~\cite{stat})   
      \begin{eqnarray}
  &\quad&    \Delta^{2}_{inf,L}\hat{x}\geq\langle
       \{\hat{x}-\langle \hat{x}\rangle-g(\hat{x}^{B}¥-\langle 
       \hat{x}^{B}¥\rangle)\}^{2}
       \rangle\nonumber\\
       &=&\sum_{x,x_{i}^{B}¥}\sum_{\lambda}¥P_{\lambda}¥\langle x|\langle 
      x_{i}^{B}¥|\rho_{\lambda}^{A}¥\rho_{\lambda}^{B}¥
      \{\hat{x}-\langle\hat{x}\rangle-g(\hat{x}^{B}¥-\langle 
      \hat{x}^{B}¥\rangle)\}^{2}
      |x\rangle |x_{i}^{B}¥\rangle \nonumber\\
         &=&\sum_{\lambda}¥P_{\lambda}¥\langle(\hat{\delta_{0}¥}-\langle 
      \hat{\delta_{0}}\rangle)^{2}
     \rangle_{\lambda}
     \geq \sum_{\lambda}P_{\lambda}
     \langle(\hat{\delta_{0}}-\langle \hat{\delta_{0}}\rangle_{\lambda}¥)^{2}
     \rangle_{\lambda}
 \end{eqnarray}
Here $\hat{\delta}_{0}¥=\hat{x}-g\hat{x}^{B}¥$ and $\langle 
\hat{q}\rangle_{\lambda}$ denotes the average  for state $\lambda$ given by density 
operator $\rho_{\lambda}=\rho_{\lambda}^{A}\rho_{\lambda}^{B}$. 
Since  $\rho_{\lambda}$ 
factorizes, 
$\langle\hat{x}\hat{x}^{B}¥\rangle_{\lambda}¥=\langle \hat{x} 
\rangle_{\lambda}¥\langle\hat{x}^{B}¥\rangle_{\lambda}¥$.  
We have 
\begin{eqnarray}
      \Delta^{2}_{inf,L}\hat{x}&\geq& \sum_{r}P_{\lambda}(\langle 
      \hat{\delta^{2}_{0}¥}\rangle_{\lambda}¥-
      \langle\hat{\delta_{0}¥}\rangle_{\lambda}^{2}¥)\nonumber\\
      &=& \sum_{\lambda}P_{\lambda}(\Delta^{2}_{\lambda}\hat{x}+g^{2}¥
     \Delta^{2}_{\lambda}\hat{x}^{B}¥)
     \end{eqnarray}
where $\Delta^{2}_{\lambda}\hat{x}=\sigma_{\lambda}^{2}(x)$ and 
$\Delta^{2}_{\lambda}\hat{x}^{B}¥=\langle 
\hat{{x}^{B}}^{2}\rangle_{\lambda}-\langle\hat{x}^{B}¥\rangle_{\lambda}^{2}$.
Also  
\begin{eqnarray}
     \Delta^{2}_{inf,L}\hat{p}
     &\geq& \sum_{\lambda}P_{\lambda}(\Delta^{2}_{\lambda}\hat{p}+h^{2}
     \Delta^{2}_{\lambda}\hat{p}^{B}¥)
     \end{eqnarray} 
     where 
     $\Delta^{2}_{\lambda}\hat{p}=\sigma_{\lambda}^{2}(p)$ and
     $\hat{p}^{B}¥$ is the 
     measurement at $B$ used to infer the result for 
     $\hat{p}$ at $A$.
 It follows (take $\Delta \hat{x}^{B}¥ \Delta \hat{p}^{B}¥\geq 1 $)
\begin{eqnarray}
 \Delta^{2}_{inf,L}\hat{x}\Delta^{2}_{inf,L}\hat{p}
       &\geq& \sum_{\lambda}P_{\lambda}¥\{ 
      \sigma_{\lambda}^{2}(x)+g^{2}\Delta^{2}_{\lambda}\hat{x}^{B}¥\} \nonumber\\
     &\quad& \quad \sum_{r}P_{r}¥ 
      \{\sigma_{\lambda}^{2}(p)+h^{2}\Delta^{2}_{\lambda}\hat{p}^{B}¥\}.
      \end{eqnarray} 
      Separability 
      implies 
      \begin{eqnarray}
 &\quad&     
 \Delta^{2}_{inf,L}\hat{x}\Delta^{2}_{inf,L}\hat{p}\nonumber\\
&\quad&\geq \langle \{ 
      \sigma_{\lambda}^{2}(x)+g^{2}\Delta^{2}_{\lambda}\hat{x}^{B}¥\}\rangle
      \langle\{\sigma_{\lambda}^{2}(p)+h^{2}\Delta^{2}_{\lambda}\hat{p}^{B}¥\}\rangle 
      \nonumber\\
&\quad& \quad\quad\geq\langle \left[
\{ 
      \sigma_{\lambda}^{2}(x)+g^{2}\Delta^{2}_{\lambda}\hat{x}^{B}¥\}
      \{\sigma_{\lambda}^{2}(p)+h^{2}\Delta^{2}_{\lambda}\hat{p}^{B}¥\}
      \right]^{1/2}
      \rangle^{2}\nonumber\\
&\quad & \quad\quad\geq\langle\{ 1+g^{2}h^{2}+
g^{2}\sigma_{\lambda}^{2}(p)\Delta^{2}_{\lambda}\hat{x}^{B}¥\nonumber\\
&\quad& \quad\quad\quad\quad\quad\quad +h^{2}\sigma_{\lambda}^{2}(x)
\Delta^{2}_{\lambda}\hat{p}^{B}¥\}^{1/2}¥\rangle^{2}¥\label{eqn:eprcondition}
\end{eqnarray}
 We notice immediately that 
 \begin{eqnarray}
      \Delta^{2}_{inf,L}\hat{x}\Delta^{2}_{inf,L}\hat{p}\geq 
     1+g^{2}h^{2}¥
\end{eqnarray}
meaning that the experimental observation of the EPR criterion
\begin{equation} 
\Delta_{inf,L}\hat{x}\Delta_{inf,L}\hat{p}< 
     1
     \label{eqn:eprlincrit}
     \end{equation}
      implies not only EPR correlations but entanglement 
     (inseparability). In terms of the quadrature amplitudes defined in 
     Section 3c, we may use this criterion to write the following 
     criterion sufficient to demonstrate not only (strong) EPR 
     correlations in the fashion of the 1935 EPR paradox, 
     but to demonstrate entanglement.
      \begin{equation}
      \Delta(X_{a}-gX_{b})\Delta(P_{a}+gP_{b})<1 
      \label{eqn:eprcritquad}
      \end{equation}

\subsection{An EPR-entanglement criterion based on the 
observation of two-mode squeezing}
    
     In fact inseparability (and the generalized weaker EPR correlations of the 
     type discussed in Section 3d) 
     may be deduced through a weaker (more easily achieved) criterion 
     based on the observation of a two-mode squeezing.
 Upon taking $g=1$, we see from 
 (\ref{eqn:eprcondition}) that another  
  constraint following necessarily from the assumption of
   separability follows. Separability implies
  \begin{eqnarray}
  &\quad&  \Delta^{2}_{inf,L}\hat{x}\Delta^{2}_{inf,L}\hat{p} 
  \nonumber\\
    &\quad& =\langle \{\hat{x}-\langle \hat{x}\rangle-(\hat{x}^{B}¥-\langle 
       \hat{x}^{B}¥\rangle )\}^{2}¥
       \rangle\langle
       \{\hat{p}-\langle \hat{p}\rangle-(\hat{p}^{B}¥-\langle 
       \hat{p}^{B}¥\rangle)\}^{2}¥
       \rangle \nonumber\\
         &\quad&\geq \langle\{ 2
\sigma_{\lambda}^{2}(p)\Delta^{2}_{\lambda}\hat{x}^{B}¥+\nonumber\\
&\quad&\quad\quad \quad\quad\quad +
1/\sigma_{\lambda}^{2}(p)\Delta^{2}_{\lambda}\hat{x}^{B}¥\}^{1/2}¥\rangle^{2}
      \geq 4   \label{eqn:sepcrit}
       \end{eqnarray}
        where we have used the inequality $x+1/x \geq 2$ and where 
        $\hat{x}^{B}$, $\hat{p}^{B}¥$ are the observables for the 
        measurements made 
       for system $B$, to allow inference of the result $\hat{x}$ 
       and $\hat{p}$ respectively at $A$. 
             The criterion (\ref{eqn:sepcrit}), following necessarily 
             from the assumption of quantum separability, then 
             implies the following criterion sufficient 
             to demonstrate entanglement (inseparability)
      \begin{eqnarray}
  &\quad&  \Delta^{2}_{inf,L}\hat{x}\Delta^{2}_{inf,L}\hat{p} 
  \nonumber\\
    &\quad& =\langle \{\hat{x}-\langle \hat{x}\rangle-(\hat{x}^{B}¥-\langle 
       \hat{x}^{B}¥\rangle )\}^{2}¥
       \rangle\langle
       \{\hat{p}-\langle \hat{p}\rangle-(\hat{p}^{B}¥-\langle 
       \hat{p}^{B}¥\rangle)\}^{2}¥
       \rangle \nonumber\\
       &\quad&    < 4   \label{eqn:insepcrit}
       \end{eqnarray}
   which may be rewritten in terms of the 
             quadrature amplitudes as 
      \begin{eqnarray}
      \Delta^{2}(X_{a}-X_{b})\Delta^{2}(P_{a}+P_{b})<4
      \end{eqnarray}
       This observation of this entanglement criterion may be 
       identified as a ``two-mode squeezing'' 
       criterion for entanglement, since the individual criterion
       \begin{eqnarray}
       \langle
       \{\hat{x}-\langle \hat{x}\rangle-(\hat{x}^{B}¥-\langle 
       \hat{x}^{B}¥\rangle)\}^{2}¥
       \rangle &<& 2
       \end{eqnarray}
     ($\Delta^{2}(X_{a}-X_{b})<2$) is the criterion for the 
     observation of a  two-mode 
     squeezing. 
      In this way we see that 
      fields that are two-mode squeezed with respect to both 
    $X_{a}-X_{b}$ and $P_{a}+P_{b}$, so as to violate (\ref{eqn:sepcrit}), 
      are necessarily entangled, and EPR correlated in the general 
      sense discussed Sections 3d. This criteria for a 
      generalized 
      demonstration of EPR correlations is generally more easily 
      achieved than that given 
      by (\ref{eqn:eprcritquad}) which is based on 
      conditional probabilities. 
      However it is pointed out that for 
      the ideal two-mode squeezed state (\ref{eqn:twomode}) both the strong 
      and the weak EPR 
      criterion are met for any nonzero value of the squeeze parameter 
      $r$.

\section{Conclusion}

In conclusion, we have established certain criteria sufficient to 
demonstrate EPR correlations. These EPR correlations are proven, not 
by violations of Bell inequalities, but  
by demonstrating a sufficient correlation between results of measurements 
performed at two spatially separated locations, where it is necessary 
to consider both ``position'' and ``momentum'' measurements. 

The EPR 
criteria are also sufficient to prove entanglement. Such 
criteria are particularly useful to situations where measurements have 
 continuous variable outcomes, where violations of Bell-type 
 inequalities are not so readily constructed. In this case 
 entanglement and EPR 
 correlations are demonstrable using highly efficient quadrature phase 
 amplitude measurements on two-mode squeezed light.
 In a second paper~\cite{mdrcrytwo} the application of strong 
 EPR criteria to give proof of 
 security in continuous variable cryptography is presented.
 
 I wish to acknowledge P. D. Drummond and S. 
 Braunstein for stimulating discussions.

 \section{Appendix}
	
	 In Section 3 the demonstration of strong EPR correlations most in 
  spirit with the original paradox was discussed. Here we require to 
  measure $\Delta_{i}x=0$, $\Delta_{j}p=0$. 
  We aim to prove as a matter of completeness 
  that any local  hidden 
  variable theory in order to predict these perfect EPR correlations 
  will involve hidden variables where the results of measurement for 
  $\hat{x}$ and $\hat{p}$ are predetermined with zero uncertainty. 
  We argue as follows.
  The assumption of a local 
  realistic theory as discussed in Sections 3d implies the local 
  hidden variable form  
  (\ref{eqn:bellsep}). The predictions for the 
  conditional distributions based on (\ref{eqn:bellsep}) are then 
  \begin{eqnarray}
   P(x|x_{i}^{B}¥)&=&\sum_{\lambda}¥P_{\lambda}¥P_{\lambda}¥(x,x_{i}^{B}¥)
   /(\sum_{\lambda}
  ¥P_{\lambda}¥P_{\lambda}(x_{i}^{B}¥))\nonumber\\
  &=&\sum_{\lambda}¥f_{\lambda}(x_{i}^{B})P_{\lambda}(x|x_{i}^{B}¥)
  \label{eqn:epropt}
  \end{eqnarray} 
  where we use a discrete summation over the possible hidden variable 
  states $\lambda$  for 
  convenience of notation only. Here $P_{\lambda}¥$ (the discrete from of 
  $\rho(\lambda)$) is the probability 
  of the system being in the hidden variable state denoted by 
  $\lambda$;  $P_{\lambda}¥(x,x_{i}^{B}¥)$ is the probability for 
  results $x$ and $x_{i}^{B}¥$ respectively upon joint measurement $\hat{x}$ and 
  $\hat{x}^{B}$ for the state $\lambda$; $P_{\lambda}(x_{i}^{B}¥)$ is the probability for result 
  $x_{i}^{B}¥$ upon measurement of $\hat{x}^{B}$, given the system is 
  in $\lambda$; and we define the fraction $f_{\lambda}(x_{i}^{B}¥)=
 P_{\lambda}¥P_{\lambda}(x_{i}^{B}¥)/(\sum_{\lambda}
  ¥P(\lambda)P_{\lambda}(x_{i}^{B}¥))$. The (\ref{eqn:epropt})
   would always imply 
  \begin{equation}
  \Delta_{i}^{2}¥x\geq \sum_{\lambda}¥f_{\lambda}(x_{i}^{B}¥)
  \Delta_{\lambda,i}^{2}¥x
 \end{equation}
 where $\Delta_{\lambda,i}^{2}¥x$ is the variance of the conditional 
 distribution $P_{\lambda}(x|x_{i}^{B}¥)$. 
 (See equations (\ref{eqn:deltaconstraint}) for a more 
 complete explanation of a similar 
 result). Recalling from (\ref{eqn:bellsep}) that 
 $P_{\lambda}(x,x_{i}^{B}¥)=P_{\lambda}(x)P_{\lambda}(x_{i}^{B}¥)$ we see 
 that $\Delta_{\lambda}x=\Delta_{\lambda,i}x$ where 
 $\Delta_{\lambda}x$ is the variance of the distribution 
 $P_{\lambda}(x)$ ($P_{\lambda}(x)$ being the probability
  that the result of $\hat{x}$ is $x$, given that the 
 system is in the hidden variable state $\lambda$). 
 Immediately then we see that where $\Delta_{i}¥x=0$, each 
 $\Delta_{\lambda}¥x=0$. This means that the system, which according 
 to the local realistic assumption is describable as being in  one 
 of the states depicted by $\lambda$ with a probability 
 $P_{\lambda}¥$, must for every $\lambda$ have a zero uncertainty 
 $\Delta_{\lambda}x$ in the prediction for result of measurement $\hat{x}$.
 The result of $\hat{x}$ is, under the local realism assumption,
  predetermined with zero uncertainty. A 
 similar conclusion is drawn for the predetermined nature of $\hat{p}$, 
 and  the EPR argument follows. 
 
  Generally the measurement of zero variances is difficult. However 
  suppose the conditional distributions $P(x|x_{i}^{B}¥)$ 
 are for each $i$  measured to be sufficiently narrow, so that there is 
 a zero probability of obtaining  a result $x$ for $\hat{x}$ at $A$ 
 which deviates from the mean $\mu_{i}$ of $P(x|x_{i}^{B})$ by 
 an amount greater than $\delta$. It follows from (\ref{eqn:epropt}) that  
 this must also  be true for each of the $P_{\lambda}¥(x|x_{i}^{B}¥)$, 
 and this in turn implies
  $\Delta_{\lambda,i}^{2}¥x\leq \delta^{2}¥$, 
  $\Delta_{\lambda,i}^{2}¥x$ being the variance of  
 $P_{\lambda}¥(x/x_{i}^{B}¥)$.  This is true {\it for 
 all} 
 $\lambda$ (and for all $i$). Recalling again that 
  $P_{\lambda}(x,x_{i}^{B}¥)=P_{\lambda}(x)P_{\lambda}(x_{i}^{B}¥)$ 
  so that $\Delta_{\lambda}x=\Delta_{\lambda,i}x$, we 
 can then conclude that the system, if compatible with local realism, must
  {\it always} be in a state where the result for $\hat{x}$ is 
 predetermined to an uncertainty $\Delta_{\lambda}x\leq\delta$ 
   where $\delta^{2}¥<1$.  
 Applying the same logic to measurements $\hat{p}$, we obtain the EPR 
 paradox.
 
  While this case of very strong EPR 
 correlation is most similar to the original EPR argument, the weaker 
 constraints described in Sections 3d and 5 still imply EPR-type correlations, in the 
 sense that it is proved that any 
 theory compatible with local realism can {\it only} predict 
 the measured correlations, if at least 
 {\it one} of the $\lambda$ ( a $\lambda_{0}¥$ say) has a sufficiently 
 definite simultaneous prediction for its result of $\hat{x}$ and $\hat{p}$
 ($\Delta_{\lambda_{0}¥}x\Delta_{\lambda_{0}¥}x\leq 1$).


\begin{references}
\bibitem{stat} This choice $d=\mu_{i}$ will minimize the rms error 
  $\sqrt{\langle (x|x_{i}^{B}¥-d)^{2}\rangle}$ where $x|x_{i}^{B}$ 
  refers to results $x$ at $A$, conditional on a result $x_{i}^{B}$ at $B$.
\bibitem{fuzzy}
  It is certainly true however that for the case of 
 perfect EPR correlations where all $\Delta _{i}x,\Delta _{j}p=0$, it 
 is required of any local realistic theory that all $\Delta_{\lambda}x= 
 \Delta_{\lambda}p=0$  
 to give hidden variables with definite  
 predictions for $\hat{x}$ and $\hat{p}$. This is proved explicitly in the 
 Appendix (as is the case of near-perfect correlation), and gives 
 the situation of the original EPR paradox discussed in Section 3a.
\bibitem{disc} 
  The degree 
   of fuzziness however cannot be determined directly from the values 
   $\Delta _{i}^{2}¥x,\Delta _{j}^{2}¥p$ for the variances of the 
   conditional distributions. This is because  poor 
   correlation between results for $A$ and $B$ 
   can be described, in a way perfectly consistent with 
   local realism, using local realistic (hidden) variables with varying definiteness 
   $\Delta_{\lambda}x$ 
 $\Delta_{\lambda}p$, 
   depending on the degree of correlation between the underlying local 
   realistic 
   variables describing the subsystems $A$ and $B$.
    For example we may have poorly-correlated  variables 
      with  
   definite predictions ($\Delta_{\lambda}x$, 
 $\Delta_{\lambda}p$ are zero), or 
   well-correlated  variables for $A$ and $B$ which have 
   with fuzzy predictions ($\Delta_{\lambda}x$, 
 $\Delta_{\lambda}p$ are large).
\end{references}
\end{document}